\documentclass[aps,prd,onecolumn,showpacs,showkeys,amsmath,amssymb]{revtex4}
\usepackage{graphicx}
\usepackage{dcolumn}
\usepackage{bm}

\begin{document}

\title{Baryons with $U_L(3)\times U_R(3)$ Chiral Symmetry IV: Interactions with Chiral
(8,1)$\oplus$(1,8) Vector and Axial-vector Mesons and Anomalous Magnetic Moments}
\author{Hua-Xing Chen$^{1,2}$}
\email{hxchen@rcnp.osaka-u.ac.jp}
\author{V. Dmitra\v sinovi\' c$^3$}
\email{dmitrasin@ipb.ac.rs}
\author{Atsushi Hosaka$^{4}$}
\email{hosaka@rcnp.osaka-u.ac.jp}
\affiliation{$^1$ School of Physics and Nuclear Energy Engineering, Beihang University, Beijing 100191, China
\\ $^2$ Departamento de F\'{\i}sica Te\'orica and IFIC, Centro Mixto Universidad de Valencia-CSIC,
Institutos de Investigaci\'on de Paterna, Aptdo. 22085, 46071 Valencia, Spain
\\ $^3$ Institute of Physics, Belgrade University, Pregrevica 118, 
11080 Beograd, Serbia \\
$^4$ Research Center for Nuclear Physics, Osaka University, Ibaraki 567--0047, Japan}

\begin{abstract}
We construct all $SU_{L}(3) \times SU_{R}(3)$ chirally invariant anomalous
magnetic, i.e. involving a Pauli tensor and one-derivative, interactions of
one chiral-$[({\bf 8},{\bf 1}) \oplus ({\bf 1}, {\bf 8})]$ meson field with
chiral-$[({\bf 6},{\bf 3})\oplus({\bf 3},{\bf 6})]$, $[({\bf 3},\overline{{\bf 3}})
\oplus (\overline{{\bf 3}}, {\bf 3})]$, and $[({\bf 8},{\bf 1}) \oplus
({\bf 1}, {\bf 8})]$ baryon fields and their ``mirror'' images.
We find strong chiral selection rules: e.g. there is only one off-diagonal
chirally symmetric anomalous magnetic interaction between $J=\frac12$
fields belonging to the $[({\bf 6},{\bf 3})\oplus({\bf 3},{\bf 6})]$ and
the $[({\bf 3}, \overline{{\bf 3}}) \oplus (\overline{{\bf 3}},{\bf 3})]$
chiral multiplets. We also study the chiral selection rules for the anomalous
magnetic interactions of the $[({\bf 3}, \overline{{\bf 3}}) \oplus
(\overline{{\bf 3}}, {\bf 3})]$ and the $[({\bf 8},{\bf 1}) \oplus
({\bf 1}, {\bf 8})]$ baryon fields. Again, no diagonal and only one off-diagonal
chiral $SU_{L}(3) \times SU_{R}(3)$ interaction of this type is allowed, that
turns out also to conserve the $U_A(1)$ symmetry.
We calculate the $F/D$ ratios for the baryons' anomalous magnetic moments
predicted by these interactions in the $SU(3)$ symmetry limit and find that only the
$[({\bf 6},{\bf 3})\oplus({\bf 3},{\bf 6})]$-$[({\bf 3}, \overline{{\bf 3}})
\oplus (\overline{{\bf 3}},{\bf 3})]$ one, reproduces $F/D$=1/3, in close
proximity to the value extracted from experiment.
\end{abstract}
%
\maketitle
\pagenumbering{arabic}
%

\section{Introduction}
\label{Intro}

Our basic assumption is that baryons are linear combinations of three
basic chiral representations ($[({\bf 6},{\bf 3})\oplus({\bf
3},{\bf 6})]$, $[({\bf 3},\overline{{\bf 3}}) \oplus
(\overline{{\bf 3}}$, and ${\bf 3})]$, $[({\bf 8},{\bf 1}) \oplus ({\bf
1}, {\bf 8})]$ \footnote{These chiral multiplets are
not limited to three-quark interpolators: for a discussion of the
validity of our assumptions, see Sect.~\ref{ssect:summary}.})
formed by three-quark interpolating fields.
Recent studies~\cite{Dmitrasinovic:2009vp,Chen:2009sf} point
towards baryon chiral mixing of $[({\bf 6},{\bf 3})\oplus({\bf
3},{\bf 6})]$ with either $[({\bf 3},\overline{{\bf 3}}) \oplus
(\overline{{\bf 3}}, {\bf 3})]$, or $[({\bf 8},{\bf 1}) \oplus ({\bf
1}, {\bf 8})]$ chiral multiplets as a possible
mechanism underlying the baryons' axial couplings.
This finding is in line with the old current algebra results
of Gerstein and Lee~\cite{Gerstein:1966zz,Gerstein:1966zy} and
of Harari~\cite{Harari:1966yq,Harari:1966jz}, updated to include
the (most recent) $F$ and $D$ values extracted from experiment
Ref.~\cite{Yamanishi:2007zza}, and extended to include the
flavor-singlet coupling $g_A^{(0)}$ of the nucleon~\cite{Bass:2007zzb,Vogelsang:2007zza},
that was not considered in the mid-1960's at all, presumably due to the lack of
data. Our own starting point were the QCD interpolating fields'
$U_A(1)$ chiral properties~\cite{Nagata:2007di,Nagata:2008zzc,Chen:2008qv}.
In Ref. \cite{Chen:2010ba} it has been shown that both the Gerstein-Lee
~\cite{Gerstein:1966zz,Gerstein:1966zy} and the Harari
~\cite{Harari:1966yq,Harari:1966jz}) scenario survive in chiral
Lagrangian models that constrain the baryon masses.

Having thus made the first step, {\it viz.} to reproduce the
phenomenological mixing starting from a chiral effective model
interaction, we turn to the next step, which is to look for
chirally symmetric dynamics that produce anomalous magnetic
moments. One such mechanism is the simplest chirally symmetric {\it
one-derivative} one-$(\rho,a)$-meson interaction Lagrangian;
one-derivative because only thus can one couple the baryon
magnetic moment (the Pauli current) to the $\rho$-field. Here we
study vector meson couplings because photon couplings follow them
under the vector meson dominance (VMD) hypothesis which has been shown
to work in the low energy region. We note, however, that the VMD
hypothesis is merely a convenient, but not necessary device, as
Gerstein and Lee~\cite{Gerstein:1966zy} have shown that the
anomalous magnetic moments of the nucleons can be obtained
using the current algebra, under the assumption that the photon
transforms as a member of the (broken chiral symmetry)
$[({\bf 8},{\bf 1}) \oplus ({\bf 1}, {\bf 8})]$ multiplet, which
amounts to VMD hypothesis with vector mesons belonging to the
$[({\bf 8},{\bf 1}) \oplus ({\bf 1}, {\bf 8})]$ chiral representation.

In this paper we construct all $SU_{L}(3) \times SU_{R}(3)$ chirally
invariant one-derivative one-vector-meson-baryon interactions and then
use them to calculate the baryons' magnetic moments. We derive the
non-derivative Dirac terms, as well. Another, perhaps equally important
and difficult problem, {\it viz.} that of the flavor-singlet anomalous
magnetic moment of the nucleon, is also addressed.
Thus our present paper serves to provide a dynamical model of chiral
mixing that is an optimal approximation to the phenomenological
solution of both the $(F, D)$ and the flavor-singlet axial
couplings, and of the anomalous magnetic moments.

In our previous publication~\cite{Chen:2010ba} we found two solutions
that fit the axial coupling data~
\footnote{this does not preclude the existence of more complicated
solutions.}: one that conserves the $U_A(1)$ symmetry (the Harari
scenario) and another one that does not (the Gerstein-Lee
scenario). Here we show that only the former scenario leads to
nucleon anomalous magnetic moments that are in agreement with
experiment extrapolated to the $SU(3)$ symmetry limit. The latter
(``Gerstein-Lee'') scenario requires vanishing nucleon anomalous magnetic
moments, in serious disagreement with experimental result extrapolated
to the $SU(3)$ symmetry limit.

Here we have, for the sake of clarity, temporarily ignored
the chiral mixing in the vector meson sector (that must violate
the $U_A(1)$-symmetry, see Ref.~\cite{Dmitrasinovic:2000ei}). This does not
affect the validity of our conclusions, as in Sect. \ref{ssect:results}
we show that the chiral interactions of vector mesons belonging to the
$[({\bf 3},\overline{{\bf 3}}) \oplus (\overline{{\bf 3}}, {\bf 3})]$
chiral multiplet with the above baryon fields lead to phenomenologically
incorrect values of the anomalous magnetic moment F/D.
We also emphasize the fact that our results (``selection rules'') hold for
arbitrary chiral mixing angles, thus making the renormalization of axial couplings
due to the axial-vector mesons irelevant for this purpose.

All this goes to show that the ``QCD $U_A(1)$ anomaly'' probably does
not play a role in the ``nucleon spin problem''~\cite{Vogelsang:2007zza,Bass:2007zzb},
as was once widely thought~\cite{Zheng:1991pk}. Rather, in all likelihood
the $U_A(1)$ anomaly provides only a (relatively) small part of the solution,
associated with the higher Fock space components, whereas the
largest part comes from the $U_A(1)$-symmetric chiral structure
of the nucleon.

In this paper we use the baryon interpolating fields to construct chirally invariant interactions. These fields have been often used in the QCD sum rule analyses~\cite{Ioffe:1981kw,Chung:1981cc,Espriu:1983hu} and Lattice QCD calculations~\cite{Zanotti:2003fx}. Most of QCD sum rule studies used the ``Ioffe current'' which leads to baryon masses consistent with those of the ground state baryons. In particular, Espriu, Pascual and Tarrach, Ref.~\cite{Espriu:1983hu} have studied the dependence on the field mixing parameter $t$ of local intepolating operators and found an optimal value around $t \approx -1$, which corresponds to the ``Ioffe current''. The Ioffe interpolating field is the same as $N_-$ in our notation, which belongs to the $[({\mathbf 3},\bar{\mathbf 3})\oplus(\bar{\mathbf 3},{\mathbf 3})]$ chiral representation~\cite{Chen:2012ex}. This is the only local interpolating field that appears in our chiral admixture results, which is consistent with the optimal choice in the QCD sum rule analyses.

This paper consists of four parts: after the present
section as introduction, in Sect.~\ref{sect:interactions} we
construct the $SU_L(3) \times SU_R(3)$ chirally invariant
interactions of non-derivative Dirac type. In Sect.~\ref{sect:anom
mag} we apply chiral mixing formalism to the hyperons'
vector-current form factors. Finally, in Sect.~\ref{sect:summary}
we discuss the results and present a summary and an outlook on the
future developments.


\section{Non-Derivative Dirac Type Interaction}
\label{sect:interactions}

In this section we propose a method for the construction of
$N_f$=3 chiral invariant vector meson-baryon interactions with
$[({\bf 8},{\bf 1}) \oplus ({\bf 1}, {\bf 8})]$ meson fields.
Both the diagonal and off-diagonal terms are possible,  which
we shall study in the following.

We have classified the baryon interpolating fields in our previous paper~\cite{Chen:2008qv}.
Our conventions for them are the same as in Refs.~\cite{Chen:2009sf,Chen:2010ba}.
Next we shall briefly define the conventions for the vector and axial-vector mesons.

\subsection{Preliminaries: Chiral Transformations of Vector and Axial-vector Mesons}
\label{ssect:su3}

We define the vector and axial-vector mesons in the $SU(3)$ space:
\begin{eqnarray}
\rho_\mu^a &=& \bar q_A \lambda_{AB}^a \gamma_\mu q_B \, ,
\\ \nonumber a_{1\mu}^a &=& \bar q_A \lambda_{AB}^a \gamma_\mu \gamma_5 q_B \, ,
\end{eqnarray}
where the index $a$ goes from 1 to 8.
They belong to the chiral representation $({\bf 8},{\bf 1}) \oplus ( {\bf 1}, {\bf 8})$.
Their linear combinations,
$M_\mu^b = \rho_\mu^b + \gamma_5 a_{1\mu}^b$ and
$M^{{\rm [mir]} b}_\mu = \gamma_0 M^{+ b}_\mu \gamma_0 = \rho_\mu^b - \gamma_5 a_{1\mu}^b$,
are the right-handed spin-one-meson vector current and the
left-handed vector current, respectively. They transform as
$({\bf 8},{\bf 1}) \oplus ( {\bf 1}, {\bf 8})$ and
$({\bf 1},{\bf 8}) \oplus ( {\bf 8}, {\bf 1})$, respectively:
\begin{eqnarray}
\delta_5^{\vec b} M_\mu^b &=& \gamma_5 b^a f_{abc} M_\mu^c
\label{def:O9} \, ,
\\ \nonumber \delta_5^{\vec b} M_\mu^{{\rm [mir]} b} &=& - \gamma_5
b^a f_{abc} M_\mu^{{\rm [mir]} c} \, .
\end{eqnarray}

To proceed our calculations sometimes we use the ``physical''
basis, whose definitions are:
\begin{eqnarray}
\left (\begin{array}{c} M_\mu^1 \\ M_\mu^2 \\ M_\mu^3 \\ M_\mu^4 \\
M_\mu^5 \\ M_\mu^6 \\ M_\mu^7 \\ M_\mu^8
\end{array} \right ) &=& \left (\begin{array}{cccccccc}
\frac{1}{\sqrt2} & -\frac{i}{\sqrt2} & 0 & 0 & 0 & 0 & 0 & 0 \\
\frac{1}{\sqrt2} & \frac{i}{\sqrt2}  & 0 & 0 & 0 & 0 & 0 & 0 \\
0 & 0 & 1 & 0 & 0 & 0 & 0 & 0 \\
0 & 0 & 0 & \frac{1}{\sqrt2} & -\frac{i}{\sqrt2} & 0 & 0 & 0 \\
0 & 0 & 0 & \frac{1}{\sqrt2} & \frac{i}{\sqrt2} & 0 & 0 & 0 \\
0 & 0 & 0 & 0 & 0 & \frac{1}{\sqrt2} & -\frac{i}{\sqrt2} & 0 \\
0 & 0 & 0 & 0 & 0 & \frac{1}{\sqrt2} & \frac{i}{\sqrt2} & 0 \\
0 & 0 & 0 & 0 & 0 & 0 & 0 & 1
\end{array} \right ) \left (\begin{array}{c}
\rho_\mu^1 + a_{1\mu}^1 \\ \rho_\mu^2 + a_{1\mu}^2 \\
\rho_\mu^3 + a_{1\mu}^3 \\ \rho_\mu^4 + a_{1\mu}^4 \\
\rho_\mu^5 + a_{1\mu}^5 \\ \rho_\mu^6 + a_{1\mu}^6 \\
\rho_\mu^7 + a_{1\mu}^7 \\ \rho_\mu^8 + a_{1\mu}^8
\end{array} \right ) \, .
\end{eqnarray}

We can define another type of $({\bf 8},{\bf 1}) \oplus ( {\bf 1}, {\bf 8})$
chiral representation: $R_\mu^b = \rho_\mu^b + a_{1\mu}^b$ as the
right-handed vector meson, and $L_\mu^b = \rho_\mu^b - a_{1\mu}^b$ as
the left-handed vector meson. They transform as
\begin{eqnarray}
\delta_5^{\vec b} R_\mu^b &=&  b^a f_{abc} R_\mu^c \label{def:D9} \, ,
\\ \nonumber \delta_5^{\vec b} L_\mu^b &=& - b^a f_{abc} L_\mu^c \, .
\end{eqnarray}
We can use these two fields to write interactions that can be used
in other calculations. We note here that these two
fields contain both the positive - and the negative parity
components, however.

\subsection{Diagonal Interactions}
\label{ssect:diag interaction}

\subsubsection{Chiral $[({\bf 6},{\bf 3})\oplus({\bf 3},{\bf 6})]$
Baryons Diagonal Interactions}
\label{ssect:(6,3)Baryons int}

To start with, it is useful to look at the chiral group structure of the
vector meson-baryon interaction $\bar N M N^\prime$, where $N$ and $N^\prime$
denote two baryon fields and $M$ denotes the vector meson fields with the Lorentz
index $\mu$ contracted either with the Dirac matrix $\gamma_\mu$ or with the
Pauli tensor $\sigma^{\mu\nu}$.

The Dirac current $\bar N \gamma_\mu N$ contains
two gamma matrices, $\gamma_\mu$ and $\gamma_0$, the latter of which comes
from the Dirac conjugate of the baryon field. Therefore, it is diagonal in
the chiral base, in other words, it takes the form:
\begin{eqnarray}
\bar N M N^\prime &\sim& \bar N_L M N^\prime_L + \bar N_R M N^\prime_R
\\ \nonumber &\sim& ( \bar N_L M_L N^\prime_L + \bar N_R M_R N^\prime_R )\,
~~~{\rm and}~~~ ( \bar N_L M^{\rm [mir]}_L N^\prime_L + \bar N_R
M^{\rm [mir]}_R N^\prime_R ) \, ,
\end{eqnarray}
when decomposed into the left and right helicity components. Then the diagonal
interaction has the structure in group representation notation
\begin{equation}\nonumber
\bar N_L(\mathbf{\bar 6}, \mathbf{\bar 3}) \times M_L(\mathbf 8, \mathbf 1)
\times N_L(\mathbf 6, \mathbf 3)
+
\bar N_R(\mathbf{\bar 3}, \mathbf{\bar 6}) \times M_R(\mathbf 1,\mathbf 8)
\times N_R(\mathbf 3, \mathbf 6) \, ,
\end{equation}
and
\begin{equation}\nonumber
\bar N_L(\mathbf{\bar 6}, \mathbf{\bar 3}) \times M_L^{\rm [mir]}(\mathbf{1},
\mathbf 8) \times N_L(\mathbf 6, \mathbf 3)
+
\bar N_R(\mathbf{\bar 3}, \mathbf{\bar 6}) \times M_R^{\rm [mir]}(\mathbf 8,
\mathbf{1}) \times N_R(\mathbf 3, \mathbf 6) \, .
\end{equation}
where in the second structure the mirror field $M_{L,R}^{\rm [mir]}$ transforms as $M_{R,L}$.

In the first term the product
$\bar N_L M_L N_L$ is decomposed as
\begin{equation}
(\mathbf{\bar 6}, \mathbf{\bar 3}) \otimes (\mathbf 8, \mathbf 1) \otimes
(\mathbf 6, \mathbf 3) \sim [(\mathbf{\bar 6}, \mathbf{\bar 3}) \otimes
(\mathbf 6, \mathbf 3)] \otimes (\mathbf 8, \mathbf 1)
\ni (\mathbf 8, \mathbf 1) \otimes (\mathbf 8, \mathbf 1) \ni (\mathbf 1, \mathbf 1) \, ,
\label{eq:63A}
\end{equation}
and the one of $\bar N_L M^{\rm [mir]}_L N_L$ is decomposed as
\begin{equation}
(\mathbf{\bar 6}, \mathbf{\bar 3}) \otimes (\mathbf 1, \mathbf 8) \otimes
(\mathbf 6, \mathbf 3) \sim [(\mathbf{\bar 6}, \mathbf{\bar 3}) \otimes
(\mathbf 6, \mathbf 3)] \otimes (\mathbf 1, \mathbf 8)
\ni (\mathbf 1, \mathbf 8) \otimes (\mathbf 1, \mathbf 8) \ni (\mathbf 1, \mathbf 1) \, .
\label{eq:63B}
\end{equation}
Therefore, there are two chiral invariant combinations for the left chirality.
The situation is the same for the right chirality. We also do this for other
diagonal and off-diagonal interactions in the following subsections.

Now we shall construct their explicit forms as
\begin{eqnarray}\nonumber
{\overline N}_{(18)}^a \gamma^\mu M^c_\mu N_{(18)}^b {\bf
C}^{abc}_{(18)} \, ,
\end{eqnarray}
and/or
\begin{eqnarray}\nonumber
{\overline N}_{(18)}^a \gamma^\mu {M}^{{\rm [mir]}c}_\mu N_{(18)}^b
{\bf C}^{abc}_{(18)} \, ,
\end{eqnarray}
where the indices $a$ and $b$ run from 1 to 18, and the index $c$
just runs from 1 to 8. By applying the chiral transformation to this
Lagrangian and demanding that this variation vanishes, we obtain
hundreds of equations, such as
\begin{eqnarray}
\delta^3_5 \big ( {\overline N}_{(18)}^a \gamma^\mu M^c_\mu N_{(18)}^b {\bf
C}^{abc}_{(18)} \big ) &=& \Big( - {\bf C}^{1,1,1}_{(18)} - \frac{2\sqrt2}{3}
{\bf C}^{1,10,1}_{(18)} + \frac{2\sqrt2}{3} {\bf C}^{10,1,1}_{(18)} \Big)
\bar p (i \gamma_5 b_3) \gamma^\mu M^1_\mu p
\\ \nonumber &+& \Big( \frac{2\sqrt2}{3} {\bf C}^{10,1,3}_{(18)} -
\frac{2\sqrt2}{3} {\bf C}^{1,10,3}_{(18)} \Big) \bar p (i \gamma_5 b_3) \gamma^\mu M^3_\mu p
\\ \nonumber &+& \cdots
\\ \nonumber &=& 0 \, .
\end{eqnarray}
Solving these equations together with the hermiticity condition, we
find that there are two solutions:
\begin{enumerate}
\item One solution can be written out using ${\bf A}_{(18)}^c$ in the
following form ($\bar N_{(18)}^a \gamma^\mu M_\mu^{c}
N_{(18)}^b {\bf C}_{(18)}^{abc}$):
\begin{equation}
\mathcal{L}^{A}_{(18)} = g^{A}_{(18)} \bar N_{(18)}^a \gamma^\mu
(\rho_\mu^c + \gamma_5 a_{1\mu}^c)
({\bf A}_{(18)}^c)_{ab} N_{(18)}^b \, ,
\label{eq:63LagA}
\end{equation}
where $g^{A}_{(18)}$ is the coupling constant, and the solution
is
\begin{eqnarray}
{\bf A}^c_{(18)} &=& \left (
\begin{array}{cc} \frac{\sqrt 3}{2}{{\bf D}_{(8)}^{c} +
{5\over2\sqrt3} {\bf F}_{(8)}^{c}} & {{\bf T}_{(8/10)}^c } \\
{{\bf T}^{\dagger c}_{(8/10)} } & {2\over\sqrt3} {\bf
F}_{(10)}^{c}
\end{array} \right ) = {\sqrt 3 \over 2} \big ( {\bf V}^c_{(18)} + {\bf F}^c_{(18)} \big ) \, .
\end{eqnarray}
The chiral group structure for this interaction is shown in Eq.~(\ref{eq:63A}).
The matrices ${\bf V}^c_{(18)}$ and ${\bf F}^c_{(18)}$
has been defined by
\begin{eqnarray}
{\bf V}^c_{(18)} =
\left(\begin{array}{cc} {\bf F}_{(8)}^{c} & 0 \\
0 & {\bf F}_{(10)}^{c}
\end{array} \right) \, , \; \; \;
\label{def:V18}
{\bf F}^c_{(18)} =
\left (
\begin{array}{cc} {{\bf D}_{(8)}^{c} +
{2 \over 3} {\bf F}_{(8)}^{c}} & {2 \over \sqrt3}{{\bf T}_{(8/10)}^c } \\
{2 \over \sqrt3} {{\bf T}^{\dagger c}_{(8/10)} } & {1\over3} {\bf
F}_{(10)}^{c}
\end{array} \right )
\end{eqnarray}
with
${\bf D}_{(8)}^{c}, {\bf F}_{(8)}^{c}, {\bf T}_{(8/10)}^c$ given in Ref.~\cite{Chen:2009sf}

\item The other solution can be written out using ${\bf B}_{(18)}^c$ in the
following form ($\bar N_{(18)}^a \gamma^\mu M_\mu^{{\rm [mir]} c}
N_{(18)}^b {\bf C}_{(18)}^{abc}$):
\begin{equation}
\mathcal{L}^{B}_{(18)} =g^{B}_{(18)} \bar N_{(18)}^a \gamma^\mu (\rho_\mu^c - \gamma_5 a_{1\mu}^c)
({\bf B}_{(18)}^c)_{ab} N_{(18)}^b \, ,
\label{eq:63LagB}
\end{equation}
where $g^{B}_{(18)}$ is the coupling constant, and the solution
is
\begin{eqnarray}
{\bf B}^c_{(18)} &=& \left (
\begin{array}{cc} {\sqrt 3 \over 2}{{\bf D}_{(8)}^{c} -
{1\over2\sqrt3} {\bf F}_{(8)}^{c}} & {{\bf T}_{(8/10)}^c } \\
{{\bf T}^{\dagger c}_{(8/10)} } & -{1\over\sqrt3} {\bf
F}_{(10)}^{c}
\end{array} \right ) = - {\sqrt 3 \over 2} \big ( {\bf V}^c_{(18)} - {\bf F}^c_{(18)} \big ) \, .
\end{eqnarray}
The chiral group structure for this interaction is shown in Eq.~(\ref{eq:63B}).

\end{enumerate}

Besides the Lagrangians (\ref{eq:63LagA}) and (\ref{eq:63LagB}), their mirror parts
\begin{eqnarray}\nonumber
g^{A}_{(18m)} \bar N_{(18m)}^a \gamma^\mu (\rho_\mu^c - \gamma_5 a_{1\mu}^c)
({\bf A}_{(18)}^c)_{ab} N_{(18m)}^b \, , ~~~ {\rm and}
~~~ g^{B}_{(18m)} \bar N_{(18m)}^a \gamma^\mu (\rho_\mu^c + \gamma_5 a_{1\mu}^c)
({\bf B}_{(18)}^c)_{ab} N_{(18m)}^b \, ,
\end{eqnarray}
are also chiral invariant. Using these solutions, and performing the
chiral transformation, we can obtain the following relations:
\begin{eqnarray}
\label{eq:18realtions}
&& - {\bf F}^{a\dagger}_{(18)} {\bf A}^b_{(18)} + {\bf A}^b_{(18)} {\bf
F}^a_{(18)} + i f_{abc} {\bf A}^c_{(18)} = 0 \, ,
\\ \nonumber && - {\bf F}^{a\dagger}_{(18)} {\bf B}^b_{(18)} + {\bf B}^b_{(18)} {\bf
F}^a_{(18)} - i f_{abc} {\bf B}^c_{(18)} = 0 \, .
\end{eqnarray}
Note that the generators ${\bf F}^a_{(18)}$ in Eq.~(\ref{def:V18})
are hermitian matrices, i.e., ${\bf F}^{a\dagger}_{(18)} = {\bf F}^a_{(18)}$.
Therefore, Eqs.~(\ref{eq:18realtions}) turn into the familiar $SU(3) \times SU(3)$
Lie algebra commutators that have already been proven in Ref.~\cite{Chen:2009sf}.
This confirms the consistency of our present calculation with that in Ref.~\cite{Chen:2009sf},
as expected.

The solution in the physical basis ($\bar N_{(18)}^a \gamma^\mu M^c_\mu
N_{(18)}^b {\bf C}^{abc}_{(18)}$) can be obtained by the
following relations:
\begin{eqnarray}\label{eq:twobasis}
\nonumber && {\bf C}^{ab3}_{(18)} = ({\bf A}^3_{(18)})_{ab} \, ,  {\bf
C}^{ab8}_{(18)} = ({\bf A}^8_{(18)})_{ab}  \, ,
\\ && {1\over\sqrt2} ({\bf C}^{ab1}_{(18)} + {\bf C}^{ab2}_{(18)}) = ({\bf
A}^1_{(18)})_{ab} \, , {i\over\sqrt2} (-{\bf C}^{ab1}_{(18)} +
{\bf C}^{ab2}_{(18)}) = ({\bf A}^2_{(18)})_{ab}  \, ,
\\ \nonumber && {1\over\sqrt2} ({\bf C}^{ab4}_{(18)} + {\bf C}^{ab5}_{(18)}) = ({\bf
A}^4_{(18)})_{ab} \, , {i\over\sqrt2} (-{\bf C}^{ab4}_{(18)} +
{\bf C}^{ab5}_{(18)}) = ({\bf A}^5_{(18)})_{ab}  \, ,
\\ \nonumber && {1\over\sqrt2} ({\bf C}^{ab6}_{(18)} + {\bf C}^{ab7}_{(18)}) = ({\bf
A}^6_{(18)})_{ab} \, , {i\over\sqrt2} (-{\bf C}^{ab6}_{(18)} +
{\bf C}^{ab7}_{(18)}) = ({\bf A}^7_{(18)})_{ab}   \, .
\end{eqnarray}

Another strategy for finding the two chiral interactions (``solutions''), is to
study the two parity-violating baryon currents interacting with the left- and right-handed
vector mesons and then to combine them to obtain the parity conserving and parity violating
Lagrangians. 
The explicit forms of interactions that we obtained by using this strategy appear
to be the most convenient ones for practical use.

To get the first solution, we use the following right-handed and left-handed currents:
\begin{enumerate}
\item The right-handed current solution can be written in the
following form:
\begin{eqnarray}
{\cal L}^{R}_{(18)} = g^{R}_{(18)} R_\mu^c {\bf R}_{\mu(18)}^c &=&
g^{R}_{(18)} (\rho_\mu^c + a_{1\mu}^c) {\overline N}_{(18)}^a
\gamma^\mu ({\bf R}_{(18)}^c)_{ab} N_{(18)}^b
\, ,
\label{def:18 intR}
\end{eqnarray}
where $g^{R}_{(18)}$ is the right-handed current coupling
constant and we have used
\begin{eqnarray}
\nonumber
{\bf R}^c_{(18)} &=&
\left({\bf V}^c_{(18)} + \gamma_5 {\bf F}^c_{(18)}\right) =
\left(\begin{array}{cc} \gamma_5 {{\bf D}_{(8)}^{c} + (1 +
{2\over3} \gamma_5) {\bf F}_{(8)}^{c}} & {2\over\sqrt3}
\gamma_5 {{\bf T}_{(8/10)}^c } \\
{2\over\sqrt3} \gamma_5 {{\bf T}^{\dagger c}_{(8/10)} } & (1 +
{1\over3}\gamma_5) {\bf F}_{(10)}^{c}
\end{array} \right) \, ,
\end{eqnarray}
and
\begin{eqnarray}
\nonumber {\bf R}^{c}_{\mu (18)} &=& \left({\bf J}^c_{\mu
(18)} + {\bf J}^c_{\mu 5 (18)}\right) = {\overline N}_{(18)}^a
\gamma^\mu \gamma_5 \left(
\begin{array}{cc} {{\bf D}_{(8)}^{c} +
({2\over3} + \gamma_5) {\bf F}_{(8)}^{c}} & {2\over\sqrt3} {{\bf T}_{(8/10)}^c } \\
{2\over\sqrt3} {{\bf T}^{\dagger c}_{(8/10)} } & ({1\over3} +
\gamma_5) {\bf F}_{(10)}^{c}
\end{array} \right)_{ab} N_{(18)}^b \, ,
\end{eqnarray}
where
\begin{eqnarray}
{\bf J}^c_{\mu(18)} &=& {\overline N}_{(18)}^a \gamma^\mu ({\bf
V}^c_{(18)})_{ab} N_{(18)}^b \, ,
\\ \nonumber {\bf J}^c_{\mu 5(18)} &=& {\overline N}_{(18)}^a
\gamma^\mu \gamma_5 ( {\bf F}^c_{(18)})_{ab} N_{(18)}^b \, .
\end{eqnarray}

\item The left-handed current solution can be written in the
following form:
\begin{eqnarray}
{\cal L}^{L}_{(18)} = g^{L}_{(18)} L_\mu^c {\bf
L}^{c}_{\mu (18)} &=& g^{L}_{(18)} (\rho_\mu^c - a_{1\mu}^c)
{\overline N}_{(18)}^a \gamma^\mu ({\bf L}_{(18)}^c)_{ab} N_{(18)}^b
\, ,
\label{def:18 intB}
\end{eqnarray}
where $g^{L}_{(18)}$ is the left-handed current coupling
constant and we have used
\begin{eqnarray}
\nonumber
{\bf L}^c_{(18)} &=& \left({\bf V}^c_{(18)} - \gamma_5 {\bf
F}^c_{(18)}\right) = \left(\begin{array}{cc} - \gamma_5 {{\bf
D}_{(8)}^{c} + (1 - {2\over3} \gamma_5) {\bf F}_{(8)}^{c}} & -
{2\over\sqrt3} \gamma_5 {{\bf T}_{(8/10)}^c } \\
- {2\over\sqrt3} \gamma_5 {{\bf T}^{\dagger c}_{(8/10)} } & (1
- {1\over3}\gamma_5) {\bf F}_{(10)}^{c}
\end{array} \right) \, ,
\end{eqnarray}
and
\begin{eqnarray}
\nonumber {\bf L}^{c}_{\mu (18)} &=& \left({\bf J}^c_{\mu
(18)} - {\bf J}^c_{\mu 5 (18)}\right) = - {\overline N}_{(18)}^a
\gamma^\mu \gamma_5 \left(
\begin{array}{cc} {{\bf D}_{(8)}^{c} +
({2\over3} - \gamma_5) {\bf F}_{(8)}^{c}} & {2\over\sqrt3} {{\bf T}_{(8/10)}^c } \\
{2\over\sqrt3} {{\bf T}^{\dagger c}_{(8/10)} } & ({1\over3} -
\gamma_5) {\bf F}_{(10)}^{c}
\end{array} \right)_{ab} N_{(18)}^b \, .
\end{eqnarray}

\end{enumerate}
These two chiral interactions generally contain both the parity-violating and
the parity-conserving parts. Their sum also contains
both of these terms, unless $g^{L}_{(18)} = \pm g^{R}_{(18)}$, when it is either
purely parity-conserving, in the case of plus sign,
\begin{eqnarray}
{\cal L}^{\rm PC}_{(18)} &=& g^{\rm PC}_{(18)} \frac12 {\overline
N}_{(18)}^a \left[(\rho_\mu^c - a_{1\mu}^c) ({\bf L}_{(18)}^{\mu~
c}) + (\rho_\mu^c + a_{1\mu}^c) ({\bf R}_{(18)}^{\mu~ c}) \right] N_{(18)}^b
\nonumber \\
&=& g^{\rm PC}_{(18)} \left[{\bm \rho}^{\mu} \cdot {\bf J}_{\mu(18)}
+ {\mathbf a}_{1}^{\mu} \cdot {\bf J}_{\mu 5(18)} \right] \, ,
\label{def:18 intPC}
\end{eqnarray}
or purely parity-violating, in the case of the minus sign.
\begin{eqnarray}
{\cal L}^{\rm PV}_{(18)} &=& g^{\rm PV}_{(18)} \frac12 {\overline
N}_{(18)}^a \left[- (\rho_\mu^c - a_{1\mu}^c) ({\bf
L}_{(18)}^{\mu~ c}) + (\rho_\mu^c
+ a_{1\mu}^c) ({\bf R}_{(18)}^{\mu~ c}) \right] N_{(18)}^b
\nonumber \\
&=& g^{\rm PV}_{(18)} \left[{\bm \rho}^{\mu} \cdot {\bf J}_{\mu
5(18)} + {\bf a}_{1}^{\mu} \cdot {\bf J}_{\mu(18)} \right] \, .
\label{def:18 intPV}
\end{eqnarray}
Thus we have obtained the first solution, Eq.~(\ref{def:18 intPC}).

To get the second solution, we can simply multiply an extra $\gamma_5$
in front of ${\bf R}_{(18)}^c$ and ${\bf L}_{(18)}^c$, and rewrite
Eqs.~(\ref{def:18 intR}) and (\ref{def:18 intB}) to be
\begin{eqnarray}
{\cal L}^{\prime R}_{(18)} = g^{\prime R}_{(18)} R_\mu^c {\bf
R}^{\prime c}_{\mu(18)} &=& g^{\prime R}_{(18)} (\rho_\mu^c +
a_{1\mu}^c) {\overline N}_{(18)}^a \gamma^\mu \gamma_5 ({\bf
R}_{(18)}^c)_{ab} N_{(18)}^b \, ,
\\ \nonumber {\cal L}^{\prime L}_{(18)} = g^{\prime L}_{(18)} L_\mu^c {\bf
L}^{\prime c}_{\mu (18)} &=& g^{\prime L}_{(18)} (\rho_\mu^c -
a_{1\mu}^c) {\overline N}_{(18)}^a \gamma^\mu \gamma_5 ({\bf
L}_{(18)}^c)_{ab} N_{(18)}^b \, .
\label{def:18_intC}
\end{eqnarray}
They are also chiral invariant. Similarly, we can use them to construct the
parity-conserving and parity-violating parts:
\begin{eqnarray}
\label{def:18B_intPC} {\cal L}^{\prime \rm PC}_{(18)} &=& g^{\prime\rm PC}_{(18)} \left[{\bm
\rho}^{\mu} \cdot {\bf J}^\prime_{\mu(18)} + {\bf a}_{1}^{\mu} \cdot
{\bf J}^\prime_{\mu5(18)} \right] \, ,
\\ \nonumber
{\cal L}^{\prime \rm PV}_{(18)} &=& g^{\prime\rm PV}_{(18)} \left[{\bm
\rho}^{\mu} \cdot {\bf J}^\prime_{\mu5(18)} + {\bf a}_{1}^{\mu}
\cdot {\bf J}^\prime_{\mu(18)} \right] \, ,
\end{eqnarray}
where
\begin{eqnarray}
{\bf J}^{\prime c}_{\mu(18)} &=& {\overline N}_{(18)}^a \gamma^\mu ({\bf
F}^c_{(18)})_{ab} N_{(18)}^b \, ,
\\ \nonumber {\bf J}^{\prime c}_{\mu5(18)} &=& {\overline N}_{(18)}^a
\gamma^\mu \gamma_5 ( {\bf V}^c_{(18)})_{ab} N_{(18)}^b \, .
\end{eqnarray}
Thus we have obtained the second solution, Eq.~(\ref{def:18B_intPC}).

Using these solutions, and performing the chiral transformation,
we can obtain the following relations:
\begin{eqnarray}
- {\bf F}^{a\dagger}_{(18)} \gamma_5 {\bf L}^b_{(18)} + {\bf
L}^b_{(18)} {\bf F}^a_{(18)} \gamma_5 - i f_{abc} {\bf L}^c_{(18)}
= 0 \, ,
\\ \nonumber - {\bf F}^{a\dagger}_{(18)}\gamma_5 {\bf R}^b_{(18)} + {\bf R}^b_{(18)} {\bf
F}^a_{(18)} \gamma_5 + i f_{abc} {\bf R}^c_{(18)} = 0 \, .
\end{eqnarray}

We can check the equivalence of these two sets of solutions, and verify the following relations:
\begin{eqnarray}
{\mathcal{L}^{\rm PC}_{(18)} \over g^{\rm PC}_{(18)} } &=& {1\over\sqrt3}\Big(
{\mathcal{L}^{A}_{(18)} \over g^{A}_{(18)} } -
{\mathcal{L}^{B}_{(18)} \over g^{B}_{(18)} } \Big) \, ,
\\ \nonumber {\mathcal{L}^{\prime \rm PC}_{(18)} \over g^{\prime \rm PC}_{(18)}}
&=& {1\over\sqrt3}\Big( {\mathcal{L}^{A}_{(18)} \over g^{A}_{(18)} } +
{\mathcal{L}^{B}_{(18)} \over g^{B}_{(18)} } \Big) \, .
\end{eqnarray}

\subsubsection{Chiral $[({\bf 3},\overline{{\bf 3}}) \oplus (\overline{{\bf 3}}, {\bf 3})]$
Baryons Diagonal Interactions}
\label{ssect:(3,3)Baryons int}

The product of the first term inside the structure ($\bar N_L M_L N_L + \bar N_R M_R N_R$)
is decomposed as
\begin{equation}
(\mathbf{\bar 3}, \mathbf{3}) \otimes (\mathbf 8, \mathbf 1) \otimes (\mathbf 3, \mathbf {\bar 3})
\sim [(\mathbf{\bar 3}, \mathbf{3}) \otimes (\mathbf 3, \mathbf {\bar 3})] \otimes
(\mathbf 8, \mathbf 1)
\ni (\mathbf 8, \mathbf 1) \otimes (\mathbf 8, \mathbf 1) \ni (\mathbf 1, \mathbf 1) \, ,
\label{eq:33A}
\end{equation}
while the one of ($\bar N_L M^{\rm [mir]}_L N_L + \bar N_R M^{\rm [mir]}_R N_R$) is decomposed as
\begin{equation}
(\mathbf{\bar 3}, \mathbf{3}) \otimes (\mathbf 1, \mathbf 8) \otimes (\mathbf 3,
\mathbf {\bar 3}) \sim [(\mathbf{\bar 3}, \mathbf{3}) \otimes (\mathbf 3, \mathbf {\bar 3})]
\otimes (\mathbf 1, \mathbf 8)
\ni (\mathbf 1, \mathbf 8) \otimes (\mathbf 1, \mathbf 8) \ni (\mathbf 1, \mathbf 1) \, .
\label{eq:33B}
\end{equation}
Therefore, there are two chiral invariant combinations. Consequently, following the
same procedures as in the previous section, we find that there are two solutions:
\begin{enumerate}

\item One solution can be written out using ${\bf A}_{(9)}^c$ in the following form
($\bar N_{(9)}^a \gamma^\mu M_\mu^{c} N_{(9)}^b {\bf C}_{(9)}^{abc}$):
\begin{equation}
\mathcal{L}^{A}_{(9)} = g^{A}_{(9)}\bar N_{(9)}^a \gamma^\mu (\rho_\mu^c + \gamma_5 a_{1\mu}^c)
({\bf A}_{(9)}^c)_{ab} N_{(9)}^b \, ,
\label{eq:33LagA}
\end{equation}
where the solution is
\begin{eqnarray}
{\bf A}^c_{(9)} &=& \left( \begin{array}{cc} 0  &
{1\over\sqrt6}{{\bf T}^c }_{(1/8)} \\
{1\over\sqrt6}{{\bf T}_{(1/8)}^{\dagger c} } &  {1 \over 2}
{\bf D}^{c}_{(8)} + {1 \over 2}{\bf F}^{c}_{(8)}
\end{array} \right)  = {1 \over 2} \big ( {\bf V}^c_{(9)} + {\bf F}^c_{(9)} \big ) \, .
\end{eqnarray}
The chiral group structure for this interaction is shown in Eq.~(\ref{eq:33A}).
The matrices ${\bf F}^c_{(9)}$ have been defined in Eq.~(\ref{def:V9}) and
\begin{eqnarray}
{\bf V}^c_{(9)} &=&
\left(\begin{array}{cc} 0 & 0 \\
0 & {\bf F}_{(8)}^{c}
\end{array} \right) \, .
\label{def:V9}
\end{eqnarray}

\item The other solution can be written out using ${\bf B}_{(9)}^c$ in the
following form ($\bar N_{(9)}^a \gamma^\mu  M_\mu^{{\rm [mir]} c} N_{(9)}^b {\bf C}_{(9)}^{abc}$):
\begin{equation}
\mathcal{L}^{B}_{(9)} = g^{B}_{(9)}\bar N_{(9)}^a \gamma^\mu (\rho_\mu^c - \gamma_5 a_{1\mu}^c)
({\bf B}_{(9)}^c)_{ab} N_{(9)}^b \, ,
\label{eq:33LagB}
\end{equation}
where the solution is
\begin{eqnarray}
{\bf B}^c_{(9)} &=& \left( \begin{array}{cc} 0  &
{1\over\sqrt6}{{\bf T}^c }_{(1/8)} \\
{1\over\sqrt6}{{\bf T}_{(1/8)}^{\dagger c} } &  {1 \over 2}
{\bf D}^{c}_{(8)} - {1 \over 2}{\bf F}^{c}_{(8)}
\end{array} \right) = - {1 \over 2} \big ( {\bf V}^c_{(9)} - {\bf F}^c_{(9)} \big ) \, .
\end{eqnarray}
The chiral group structure of this interaction is shown in Eq.~(\ref{eq:33B}).

\end{enumerate}

Besides the Lagrangians (\ref{eq:33LagA}) and (\ref{eq:33LagB}), their mirror parts
\begin{eqnarray}\nonumber
g^{A}_{(9m)} \bar N_{(9m)}^a \gamma^\mu (\rho_\mu^c - \gamma_5 a_{1\mu}^c)
({\bf A}_{(9)}^c)_{ab} N_{(9m)}^b \, , ~~~ {\rm and} ~~~ g^{B}_{(9m)} \bar N_{(9m)}^a
\gamma^\mu (\rho_\mu^c + \gamma_5 a_{1\mu}^c)
({\bf B}_{(9)}^c)_{ab} N_{(9m)}^b \, ,
\end{eqnarray}
are also chiral invariant. Using these solutions, and performing the chiral transformation,
we can obtain the following relations:
\begin{eqnarray}\label{eq:9relations}
&& - {\bf F}^{a\dagger}_{(9)} {\bf A}^b_{(9)} + {\bf A}^b_{(9)} {\bf
F}^a_{(9)} + i f_{abc} {\bf A}^c_{(9)} = 0 \, ,
\\ \nonumber && - {\bf F}^{a\dagger}_{(9)} {\bf B}^b_{(9)} + {\bf B}^b_{(9)} {\bf
F}^a_{(9)} - i f_{abc} {\bf B}^c_{(9)} = 0 \, .
\end{eqnarray}
Note that the generators ${\bf F}^a_{(9)}$ defined in Ref. \cite{Chen:2009sf} are
hermitian matrices, i.e., ${\bf F}^{a\dagger}_{(9)} = {\bf
F}^a_{(9)}$. Therefore, Eqs. (\ref{eq:9relations}) turn into the familiar
$SU(3) \times SU(3)$ Lie commutators that have already been proven in Ref.~\cite{Chen:2009sf}.
This, once again, proves the consistency of our present calculation with that in Ref.~\cite{Chen:2009sf}.

We can use the other strategy which has been discussed in the previous section to
obtain these two interactions. One solution is
\begin{eqnarray}
\label{def:9intPC}
{\cal L}^{\rm PC}_{(9)}
&=& g^{\rm PC}_{(9)} \left[{\bm \rho}^{\mu} \cdot {\bf J}_{\mu(9)} +
{\bf a}_{1}^{\mu} \cdot {\bf J}_{\mu 5(9)} \right] \, ,
\end{eqnarray}
where
\begin{eqnarray}
{\bf J}^c_{\mu(9)} &=& {\overline N}_{(9)}^a \gamma^\mu ({\bf
V}^c_{(9)})_{ab} N_{(9)}^b \, ,
\\ \nonumber {\bf J}^c_{\mu 5(9)} &=& {\overline N}_{(9)}^a
\gamma^\mu \gamma_5 ( {\bf F}^c_{(9)})_{ab} N_{(9)}^b \, ,
\end{eqnarray}
and the other solution is (similarly obtained by adding an extra $\gamma_5$):
\begin{eqnarray}
\label{def:9B_intPC}
{\cal L}^{\prime \rm PC}_{(9)} &=& g^{\prime\rm PC}_{(9)} \left[{\bm
\rho}^{\mu} \cdot {\bf J}^\prime_{\mu(9)} + {\bf a}_{1}^{\mu} \cdot
{\bf J}^\prime_{\mu5(9)} \right] \, ,
\end{eqnarray}
where
\begin{eqnarray}
{\bf J}^{\prime c}_{\mu(9)} &=& {\overline N}_{(9)}^a \gamma^\mu ({\bf
F}^c_{(9)})_{ab} N_{(9)}^b \, ,
\\ \nonumber {\bf J}^{\prime c}_{\mu5(9)} &=& {\overline N}_{(9)}^a
\gamma^\mu \gamma_5 ( {\bf V}^c_{(9)})_{ab} N_{(9)}^b \, .
\end{eqnarray}
The relevant purely parity-violating partners are
\begin{eqnarray}
{\cal L}^{\rm PV}_{(9)}
&=& g^{\rm PV}_{(9)} \left[{\bm \rho}^{\mu} \cdot {\bf J}_{\mu 5(9)}
+ {\bf a}_{1}^{\mu} \cdot {\bf J}_{\mu(9)} \right] \, .
\label{def:9intPV}
\\ \nonumber
{\cal L}^{\prime \rm PV}_{(9)} &=& g^{\prime\rm PV}_{(9)} \left[{\bm
\rho}^{\mu} \cdot {\bf J}^\prime_{\mu5(9)} + {\bf a}_{1}^{\mu} \cdot
{\bf J}^\prime_{\mu(9)} \right] \, ,
\end{eqnarray}

We can check the equivalence of these two sets of solutions, and verify the following relations:
\begin{eqnarray}
{\mathcal{L}^{\rm PC}_{(9)} \over g^{\rm PC}_{(9)} } &=&  {\mathcal{L}^{A}_{(9)}
\over g^{A}_{(9)} } - {\mathcal{L}^{B}_{(9)} \over g^{B}_{(9)} }  \, ,
\\ \nonumber {\mathcal{L}^{\prime \rm PC}_{(9)} \over g^{\prime \rm PC}_{(9)} } &=&
{\mathcal{L}^{A}_{(9)} \over g^{A}_{(9)} } + {\mathcal{L}^{B}_{(9)} \over g^{B}_{(9)} } \, .
\end{eqnarray}

\subsubsection{Chiral $[({\bf 8},{\bf 1})
\oplus ({\bf 1}, {\bf 8})]$ Baryons Diagonal Interactions}
\label{ssect:(8,1)Baryons int}

The product of the first term inside the structure ($\bar N_L M_L N_L + \bar N_R M_R N_R$)
is decomposed as
\begin{equation}
(\mathbf{8}, \mathbf{1}) \otimes (\mathbf 8, \mathbf 1) \otimes (\mathbf 8, \mathbf {1})
\ni [(\mathbf 8, \mathbf 1) \oplus (\mathbf 8, \mathbf 1)] \otimes (\mathbf 8, \mathbf 1)
\ni (\mathbf 1, \mathbf 1) \oplus (\mathbf 1, \mathbf 1) \, ,
\label{eq:8}
\end{equation}
Therefore, there are two chiral invariant combinations.
Consequently, following the same procedure as in the previous section(s), we find
that there are two solutions. They can be written out using ${\bf
F}_{(8)}^c$ and ${\bf D}_{(8)}^c$ in the following form ($\bar N_{(8)}^a \gamma^\mu M_\mu^{c}
N_{(8)}^b {\bf C}_{(8)}^{abc}$):
\begin{eqnarray}
\mathcal{L}^F_{(8)} &=& g^F_{(8)}\bar N_{(8)}^a \gamma^\mu (\rho_\mu^c + \gamma_5 a_{1\mu}^c)
({\bf F}_{(8)}^c)_{ab} N_{(8)}^b \, ,
\label{eq:8LagA}
\\ \mathcal{L}^D_{(8)} &=& g^D_{(8)}\bar N_{(8)}^a \gamma^\mu (\rho_\mu^c + \gamma_5 a_{1\mu}^c)
({\bf D}_{(8)}^c)_{ab} N_{(8)}^b \, .
\label{eq:8LagB}
\end{eqnarray}
The chiral group structure for these two interactions is just shown in Eq.~(\ref{eq:8}).

Besides the Lagrangians (\ref{eq:8LagA}) and (\ref{eq:8LagB}), their mirror parts
\begin{eqnarray}\nonumber
g^{F}_{(8m)} \bar N_{(8m)}^a \gamma^\mu (\rho_\mu^c - \gamma_5 a_{1\mu}^c)
({\bf F}_{(8)}^c)_{ab} N_{(8m)}^b \, , ~~~ {\rm and} ~~~ g^{D}_{(8m)} \bar N_{(8m)}^a
\gamma^\mu (\rho_\mu^c - \gamma_5 a_{1\mu}^c)
({\bf D}_{(8)}^c)_{ab} N_{(8m)}^b \, ,
\end{eqnarray}
are also chiral invariant. Using these solutions, and performing the chiral transformation,
we can obtain the following relations:
\begin{eqnarray}
\label{eq:8relation}&& - {\bf F}^{a\dagger}_{(8)} {\bf F}^b_{(8)} + {\bf F}^b_{(8)} {\bf
F}^a_{(8)} + i f_{abc} {\bf F}^c_{(8)} = 0 \, , \\ \nonumber && - {\bf F}^{a\dagger}_{(8)}
{\bf D}^b_{(8)} + {\bf D}^b_{(8)} {\bf
F}^a_{(8)} + i f_{abc} {\bf D}^c_{(8)} = 0 \, .
\end{eqnarray}
Note that the generators ${\bf F}^a_{(8)}$ defined in Ref. \cite{Chen:2009sf} are
hermitian matrices, i.e., ${\bf F}^{a\dagger}_{(8)} = {\bf
F}^a_{(8)}$. Therefore, Eqs. (\ref{eq:8relation}) turn into the familiar $SU(3) \times SU(3)$ Lie
commutators that have already been proven in Ref.~\cite{Chen:2009sf}. This, once again,
proves the consistency of our present calculation with that in Ref.~\cite{Chen:2009sf}.

We can use the other strategy which has been discussed in the previous section(s) to obtain
these two interactions. One solution is
\begin{eqnarray}
{\cal L}^{\rm PC}_{(8)}
&=& g^{\rm PC}_{(8)} \left[{\bm \rho}^{\mu} \cdot {\bf J}_{\mu(8)} +
{\bf a}_{1}^{\mu} \cdot {\bf J}_{\mu 5(8)} \right] \, ,
\label{def:8intPC}
\end{eqnarray}
where
\begin{eqnarray}
{\bf J}^c_{\mu(8)} &=& {\overline N}_{(8)}^a \gamma^\mu ({\bf
F}^c_{(8)})_{ab} N_{(8)}^b \, ,
\\ \nonumber {\bf J}^c_{\mu 5(8)} &=& {\overline N}_{(8)}^a
\gamma^\mu \gamma_5 ( {\bf F}^c_{(8)})_{ab} N_{(8)}^b \, ,
\end{eqnarray}
Here, we might think that the other solution can be obtained similarly by adding
an extra $\gamma_5$ which we have done in the previous section(s). However, we
find that the solution obtained in this way is same as the original solution.
Therefore, in order to get the second solution we need to find another different
set of ${\bf J}^{\prime c}_{\mu(8)}$ and ${\bf J}^{\prime c}_{\mu 5(8)}$:
\begin{eqnarray}
{\bf J}^{\prime c}_{\mu(8)} &=& {\overline N}_{(8)}^a \gamma^\mu ({\bf
D}^c_{(8)})_{ab} N_{(8)}^b \, ,
\\ \nonumber {\bf J}^{\prime c}_{\mu 5(8)} &=& {\overline N}_{(8)}^a
\gamma^\mu \gamma_5 ( {\bf D}^c_{(8)})_{ab} N_{(8)}^b \, ,
\end{eqnarray}
and the second solution is:
\begin{eqnarray}
{\cal L}^{\prime \rm PC}_{(8)}
&=& g^{\prime \rm PC}_{(8)} \left[{\bm \rho}^{\mu} \cdot {\bf
J}^{\prime}_{\mu(8)} + {\bf a}_{1}^{\mu} \cdot {\bf J}^{\prime}_{\mu
5(8)} \right] \, . \label{def:8BintPC}
\end{eqnarray}
The relevant parity-violating partners are
\begin{eqnarray}
{\cal L}^{\rm PV}_{(8)}
&=& g^{\rm PV}_{(8)} \left[{\bm \rho}^{\mu} \cdot {\bf J}_{\mu 5(8)}
+ {\bf a}_{1}^{\mu} \cdot {\bf J}_{\mu(8)} \right] \, ,
\label{def:8intPV}
\\ {\cal L}^{\prime \rm PV}_{(8)}
&=& g^{\prime \rm PV}_{(8)} \left[{\bm \rho}^{\mu} \cdot {\bf
J}^\prime_{\mu 5(8)} + {\bf a}_{1}^{\mu} \cdot {\bf
J}^\prime_{\mu(8)} \right] \, . \label{def:8BintPV}
\end{eqnarray}

We can check the equivalence of these two sets of solutions, and verify the following relations:
\begin{eqnarray}
{\mathcal{L}^{\rm PC}_{(8)} \over g^{\rm PC}_{(8)} } &=&
{\mathcal{L}^{F}_{(8)} \over g^{F}_{(8)} } \, ,
\\ \nonumber {\mathcal{L}^{\prime \rm PC}_{(8)} \over g^{\prime \rm PC}_{(8)} } &=&
{\mathcal{L}^{D}_{(8)} \over g^{D}_{(8)} } \, .
\end{eqnarray}

\subsubsection{Chiral $[({\bf 10},{\bf 1})
\oplus ({\bf 1}, {\bf 10})]$ Baryons Diagonal Interactions}
\label{ssect:(10,1)Baryons int}

The product of the first term inside the structure ($\bar N_L M_L N_L + \bar N_R M_R N_R$)
is decomposed as
\begin{equation}
(\mathbf{\bar {10}}, \mathbf{1}) \otimes (\mathbf 8, \mathbf 1) \otimes (\mathbf {10},
\mathbf {1}) \sim [(\mathbf{\bar {10}}, \mathbf{1}) \otimes (\mathbf {10}, \mathbf {1})]
\otimes (\mathbf 8, \mathbf 1)
\ni (\mathbf 8, \mathbf 1) \otimes (\mathbf 8, \mathbf 1) \ni (\mathbf 1, \mathbf 1) \, ,
\label{eq:10}
\end{equation}
Therefore, there is only one chiral invariant combination.
Consequently, following the same procedure as in the previous section(s), we
find that there is only one solution, which can be written out
using ${\bf F}_{(10)}^c$ in the following form ($\bar \Delta_{(10)}^a \gamma^\mu M_\mu^{c}
\Delta_{(10)}^b {\bf C}_{(10)}^{abc}$):
\begin{equation}
\mathcal{L}_{(10)} = g_{(10)}\bar \Delta_{(10)}^a \gamma^\mu (\rho_\mu^c + \gamma_5 a_{1\mu}^c)
({\bf F}_{(10)}^c)_{ab} \Delta_{(10)}^b \, .
\label{eq:10Lag}
\end{equation}
The chiral group structure for these two interactions is just shown in Eq.~(\ref{eq:10}).

Besides the Lagrangian Eq. (\ref{eq:10Lag}), its mirror part
\begin{eqnarray}\nonumber
g_{(10m)} \bar \Delta_{(10m)}^a \gamma^\mu (\rho_\mu^c - \gamma_5 a_{1\mu}^c)
({\bf F}_{(10)}^c)_{ab} \Delta_{(10m)}^b \, ,
\end{eqnarray}
is also chiral invariant. Using these solutions, and performing the chiral transformation,
we can obtain the following relations:
\begin{eqnarray}
- {\bf F}^{a\dagger}_{(10)} {\bf F}^b_{(10)} + {\bf F}^b_{(10)} {\bf
F}^a_{(10)} + i f_{abc} {\bf F}^c_{(10)} = 0 \, .
\label{eq:10relation}
\end{eqnarray}
Note that the generators ${\bf F}^a_{(10)}$ defined in 
Ref. \cite{Chen:2009sf} are hermitian matrices, i.e., ${\bf F}^{a\dagger}_{(10)} = {\bf
F}^a_{(10)}$. Therefore, Eq.~(\ref{eq:10relation}) turns into the familiar $SU(3) \times SU(3)$
Lie commutators that have already been proven in Ref.~\cite{Chen:2009sf}. This,
once again, proves the consistency of our present calculation with that in Ref.~\cite{Chen:2009sf}.

We can also use the other strategy which has been discussed in the previous section(s)
to obtain this interaction:
\begin{eqnarray}
{\cal L}^{\rm PC}_{(10)}
&=& g^{\rm PC}_{(10)} \left[{\bm \rho}^{\mu} \cdot {\bf J}_{\mu(10)} +
{\bf a}_{1}^{\mu} \cdot {\bf J}_{\mu 5(10)} \right] \, ,
\label{def:10intPC}
\end{eqnarray}
where
\begin{eqnarray}
{\bf J}^c_{\mu(10)} &=& {\overline N}_{(10)}^a \gamma^\mu ({\bf
F}^c_{(10)})_{ab} N_{(10)}^b \, ,
\\ \nonumber {\bf J}^c_{\mu 5(10)} &=& {\overline N}_{(10)}^a
\gamma^\mu \gamma_5 ( {\bf F}^c_{(10)})_{ab} N_{(10)}^b \, .
\end{eqnarray}
The relevant parity-violating partner is
\begin{eqnarray}
{\cal L}^{\rm PV}_{(10)}
&=& g^{\rm PV}_{(10)} \left[{\bm \rho}^{\mu} \cdot {\bf J}_{\mu 5(10)}
+ {\bf a}_{1}^{\mu} \cdot {\bf J}_{\mu(10)} \right] \, .
\label{def:10intPV}
\end{eqnarray}

\subsection{Chiral Mixing Interactions}
\label{ssect:off diag interaction}

To construct chiral invariant  off-diagonal interactions, we need to consider
the following off-diagonal terms in the chiral base, in other words, it can also take the form
\begin{eqnarray}
\bar N M N^\prime &\sim& \bar N_L M N^\prime_R + \bar N_R M N^\prime_L
\\ \nonumber &=& ( \bar N_L M_L N^\prime_R + \bar N_R M_R N^\prime_L ) + (\bar N_L M_R N^\prime_R + \bar N_R M_L N^\prime_L ) \, ,
\end{eqnarray}
when decomposed into the left and right components. However, to arrive at this form we need to use the mirror field $N^{\prime {\rm [mir]}}$ to have the correct helicity structure:
\begin{equation}
\bar N_L M N^\prime_R + \bar N_R M N^\prime_L \sim \bar N_L M N^{\prime \rm [mir]}_L + \bar N_R M N^{\prime \rm [mir]}_R \, .
\end{equation}

\subsubsection{Chiral Mixing Interaction
$[({\bf 6},{\bf 3})\oplus({\bf 3},{\bf 6})]$ - $[({\bf 3},
\overline{\bf 3}) \oplus (\overline{\bf 3}, {\bf 3})]$}
\label{ssect:(6,3)(3,3)interaction}

The product of the first term inside ($\bar N_L M_L N^{\prime \rm [mir]}_L + \bar N_R M_R N^{\prime \rm [mir]}_R$) is decomposed as
\begin{equation}
(\mathbf{\bar {6}}, \mathbf{\bar 3}) \otimes (\mathbf 8, \mathbf 1) \otimes (\mathbf {\bar 3}, \mathbf {3}) \sim [(\mathbf{\bar {6}}, \mathbf{\bar 3}) \otimes (\mathbf {\bar 3}, \mathbf {3})] \otimes (\mathbf 8, \mathbf 1)
\ni (\mathbf 8, \mathbf 1) \otimes (\mathbf 8, \mathbf 1) \ni (\mathbf 1, \mathbf 1) \, ,
\end{equation}
Therefore, there is one chiral invariant combination, and
we find that the mixing of $[({\bf 6},{\bf 3})\oplus({\bf 3},{\bf
6})]$ with $[(\overline{\bf 3}, {\bf 3}) \oplus ({\bf 3},
\overline{\bf 3})]_{\rm [mir]}$ baryon fields together with an
$[(\bf 8,\bf 1)\oplus(\bf 1,\bf 8)]$ chiral multiplet of vector
and axial-vector meson fields can form a chiral singlet. We find
the following form of the chiral invariant interaction
\begin{eqnarray}
\label{e:9/18_Dirac_interact}
{\overline N}^a_{(9m)} \gamma^\mu M^{c}_\mu N_{(18)}^b {\bf C}^{abc}_{(9/18)} + h.c. \, .
\end{eqnarray}
The coefficients ${\bf C}^{abc}_{(9/18)}$ can be similarly
obtained as in Eq.~(\ref{eq:twobasis}), and once again we find a
parity-conserving interaction
\begin{eqnarray}
{\cal L}^{\rm PC}_{(9/18)} &=& g^{\rm PC}_{(9/18)} \left[ {\overline
N}^a_{(9m)} \gamma^\mu (\rho_\mu^c + \gamma_5 a_{1\mu}^c) ({\bf
T}^c_{(9/18)})_{ab} N^b_{(18)} + h.c.\right] \label{e:9/18 PC Dirac_interact} \, ,
\end{eqnarray}
and a parity-violating partner
\begin{eqnarray}
{\cal L}^{\rm PV}_{(9/18)} &=& g^{\rm PV}_{(9/18)} \left[ {\overline
N}^a_{(9m)} \gamma^\mu \gamma_5 (\rho_\mu^c + \gamma_5 a_{1\mu}^c) ({\bf
T}^c_{(9/18)})_{ab} N^b_{(18)} + h.c.\right] \label{e:9/18_PV Dirac_interact} \, ,
\end{eqnarray}
with
\begin{eqnarray}
{\bf T}^c_{(9/18)} &=& \left ( \begin{array}{cc}
{1\over2} {\bf T}^c_{(1/8)} & {\bf 0}_{1\times10} \\
-{\sqrt3\over2\sqrt2}{\bf D}^c_{(8)}-{1\over2\sqrt6}{\bf F}^c_{(8)}
& {1\over\sqrt2} {\bf T}^c_{(8/10)}
\end{array} \right )
\label{e:9/18 matrix} \, ,
\end{eqnarray}
that satisfies the following relation:
\begin{eqnarray}
{\bf F}^{a\dagger}_{(9)} {\bf T}^b_{(9/18)} + {\bf T}^b_{(9/18)}
{\bf F}^a_{(18)} + i f_{abc} {\bf T}^c_{(9/18)} = 0 \, .
\end{eqnarray}

\subsubsection{Chiral Mixing Interaction
$[({\bf 10},{\bf 1})\oplus({\bf 1},{\bf 10})]$ -- $[({\bf 8},{\bf
1})\oplus({\bf 1},{\bf 8})]$}
\label{ssect:(10,1)(8,1)interaction}

The product of the first term inside ($\bar N_L M_L N_L + \bar N_R M_R N_R$) is decomposed as
\begin{equation}
(\mathbf{\bar {10}}, \mathbf{1}) \otimes (\mathbf 8, \mathbf 1) \otimes (\mathbf {8},
\mathbf {1}) \sim [(\mathbf{\bar {10}}, \mathbf{1}) \otimes (\mathbf {8}, \mathbf {1})]
\otimes (\mathbf 8, \mathbf 1) \ni (\mathbf 8, \mathbf 1) \otimes (\mathbf 8, \mathbf 1)
\ni (\mathbf 1, \mathbf 1) \, ,
\end{equation}
Therefore, there is one chiral invariant combination, and
we find that the mixing of $[({\bf 10},{\bf 1})\oplus({\bf 1},{\bf
10})]$  with $[({\bf 8},{\bf 1})\oplus({\bf 1},{\bf 8})]$ baryon
fields together with an $[(\bf 8,\bf 1)\oplus(\bf 1,\bf 8)]$
chiral multiplet of vector and axial-vector meson fields can form
a chiral singlet. We find the following form of the chiral
invariant interaction
\begin{eqnarray}
{\overline N}^a_{(8)} \gamma^\mu M^{c}_\mu N_{(10)}^b
{\bf C}^{abc}_{(8/18)} + h.c. \, .
\end{eqnarray}
The coefficients ${\bf C}^{abc}_{(8/18)}$ can be similarly
obtained as in Eq.~(\ref{eq:twobasis}), and once again we find a
parity-conserving interaction
\begin{eqnarray}
{\cal L}^{\rm PC}_{(8/18)} &=& g^{\rm PC}_{(8/18)} \left[ {\overline
N}^a_{(8)} \gamma^\mu (\rho_\mu^c + \gamma_5 a_{1\mu}^c) ({\bf
T}^c_{(8/10)})_{ab} N^b_{(10)} + h.c.\right] \label{e:8/10 PC Dirac_interact} \, ,
\end{eqnarray}
and a parity-violating partner
\begin{eqnarray}
{\cal L}^{\rm PV}_{(8/18)} &=& g^{\rm PV}_{(8/10)} \left[ {\overline
N}^a_{(8)} \gamma^\mu \gamma_5 (\rho_\mu^c + \gamma_5 a_{1\mu}^c) ({\bf
T}^c_{(8/10)})_{ab} N^b_{(10)} + h.c.\right] \label{e:8/10_PV Dirac_interact} \, .
\end{eqnarray}
We find that the only solution is formed by the ${\bf T}^c_{(8/10)}$
matrices defined in Ref. \cite{Chen:2009sf}
that satisfy the following relation:
\begin{eqnarray}
-{\bf F}^{a\dagger}_{(8)} {\bf T}^b_{(8/10)} + {\bf T}^b_{(8/10)}
{\bf F}^a_{(10)} + i f_{abc} {\bf T}^c_{(8/10)} = 0 \, .
\end{eqnarray}

\subsection{Brief Summary of the Dirac type Interactions}
\label{ssect:summary}

Note that all of the diagonal Dirac type interactions that were
shown here also appear in a {\it local} $SU_{L}(3) \times
SU_{R}(3)$ chiral symmetry Yang-Mills interaction. We have found
more, however: there are chiral off-diagonal interaction terms
that cannot be obtained by a minimal substitution in the kinetic
energy, of the Yang-Mills type because {\it per definitionem} the
kinetic energies are diagonal operators in the chiral
representation space. These off-diagonal terms show up in the
flavor-decimet channel, and therefore are physically less
accessible than the diagonal flavor-octet ones, which are easier
to access by way of elastic scattering. Here, as well, there are
strong chiral selection rules.

\begin{table}[hbt]
\caption{Allowed chiral invariant Dirac type interaction terms
with one $(\mathbf{8},\mathbf{1})\oplus(\mathbf{1},\mathbf{8})$
vector meson field $\bar N \gamma^\mu M_\mu N$. In the first column
we show the chiral representation of $N$, and the first row the
chiral representation of $\bar N$. We use ``[mir]'' to denote the
relevant mirror fields.}
\begin{center}
\label{tab:Dirac interactions}
\begin{tabular}{c|c|c|c|c}
\hline \hline &
$(\mathbf{8},\mathbf{1})\oplus(\mathbf{1},\mathbf{8})$ &
$(\mathbf{3},\mathbf{\bar 3})\oplus(\mathbf{\bar 3},\mathbf{3})$ &
$(\mathbf{6},\mathbf{3}) \oplus (\mathbf{3},\mathbf{6})$ &
$(\mathbf{10},\mathbf{1})\oplus(\mathbf{1},\mathbf{10})$ \\
\hline $(\mathbf{8},\mathbf{1})\oplus(\mathbf{1},\mathbf{8})$ &
2 $\times M_\mu$ &  &  & $M_\mu$ \\
\hline $(\mathbf{\bar 3},\mathbf{3})\oplus(\mathbf{3},
\mathbf{\bar 3})$ & & $M_\mu$, $M^\dagger_\mu$ & & \\
\hline $(\mathbf{\bar 6},\mathbf{\bar 3}) \oplus (\mathbf{\bar
3},\mathbf{\bar 6})$ & & & $M_\mu$, $M^\dagger_\mu$ &
\\ \hline $(\mathbf{\overline{10}},\mathbf{1})\oplus(\mathbf{1},
\mathbf{\overline{10}})$& $M_\mu$ & & & $M_\mu$
\\ \hline
\hline & $(\mathbf{3},\mathbf{\bar 3})\oplus(\mathbf{\bar 3},\mathbf{3})$
& $(\mathbf{6},\mathbf{3}) \oplus (\mathbf{3},\mathbf{6})$
\\ \cline{1-3} $(\mathbf{3},\mathbf{\bar
3})\oplus(\mathbf{\bar 3},\mathbf{3})$[mir] & & $M_\mu$  \\
\cline{1-3} $(\mathbf{\bar 3},\mathbf{\bar 6}) \oplus
(\mathbf{\bar 6},\mathbf{\bar 3})$[mir] & $M^\dagger_\mu$ &
\\ \hline \hline
\end{tabular}
\end{center}
\end{table}

\section{One-Derivative Pauli Type Interactions}
\label{sect:anom mag}

Now let us look at one-derivative Pauli type interactions. They lead
to the anomalous  magnetic moments through vector meson dominance.
The interaction terms take in general the following form:
\begin{eqnarray}
\bar N^a \sigma^{\mu\nu} \partial_\nu M_\mu^c N^b {\bf C}_{abc} \, ,
\end{eqnarray}
which has the helicity flip structure, i.e., $\bar N_L \mathcal{O} N_R$.
Due to this structure, the chiral selection rules are far more restrictive than
otherwise.
As first noted by Dashen and Gell-Mann \cite{Dashen:1965ik}, all of the diagonal
anomalous magnetic interactions must vanish due to such chiral symmetry restrictions.

All of the off-diagonal anomalous magnetic interactions can be
easily obtained from the off-diagonal and diagonal Dirac type
ones in Sect. \ref{sect:interactions} in many cases by simply
substituting one of the baryon fields with its mirror one.
Mixing of various combinations of chiral multiplets
gives the following chiral invariant interactions:

\begin{enumerate}

\item
For $[({\bf 6},{\bf 3})\oplus({\bf 3},{\bf 6})]$ -
$[(\mathbf{3},\mathbf{6}) \oplus (\mathbf{6},\mathbf{3})]_{\rm
[mir]}$ the chiral invariant interactions are:
\begin{eqnarray}
&& \left(\frac{\kappa^A_{(18)}}{2M} \right)
\overline N^a_{(18)} \sigma^{\mu\nu} \partial_\nu (\rho_\mu^c -
\gamma_5 a_{1\mu}^c) ({\bf V}^c_{(18)} + {\bf F}^c_{(18)})_{ab}
N^b_{(18m)} + h.c. \, ,
\\ \nonumber && \left(\frac{\kappa^B_{(18)}}{2M} \right)
\overline N^a_{(18)} \sigma^{\mu\nu} \partial_\nu (\rho_\mu^c +
\gamma_5 a_{1\mu}^c) ({\bf V}^c_{(18)} - {\bf F}^c_{(18)})_{ab}
N^b_{(18m)} + h.c. \, . \label{e:18/18m_Pauli_interact}
\end{eqnarray}

\item
For $[({\bf 3}, \overline{\bf 3}) \oplus (\overline{\bf 3}, {\bf 3})]$
- $[(\mathbf{\bar 3},\mathbf{3}) \oplus (\mathbf{3},\mathbf{\bar
3})]_{\rm [mir]}$ the chiral invariant interactions are:
\begin{eqnarray}
&& \left(\frac{\kappa^A_{(9)}}{2M} \right) \overline N^a_{(9)}
\sigma^{\mu\nu} \partial_\nu (\rho_\mu^c - \gamma_5 a_{1\mu}^c)
({\bf V}^c_{(9)} + {\bf F}^c_{(9)})_{ab} N^b_{(9m)} + h.c. \, ,
\\ \nonumber && \left(\frac{\kappa^B_{(9)}}{2M} \right) \overline N^a_{(9)}
\sigma^{\mu\nu} \partial_\nu (\rho_\mu^c + \gamma_5 a_{1\mu}^c)
({\bf V}^c_{(9)} - {\bf F}^c_{(9)})_{ab} N^b_{(9m)} + h.c. \, .
\label{e:9/9m_Pauli_interact}
\end{eqnarray}

\item
For $[(\bf 8,\bf 1)\oplus(\bf 1,\bf 8)]$ - $[(\bf 1,\bf 8)\oplus(\bf
8,\bf 1)]_{\rm [mir]}$  the chiral invariant interactions are:
\begin{eqnarray}
&& \left(\frac{\kappa^A_{(8)}}{2M} \right) \overline N^a_{(8)}
\sigma^{\mu\nu} \partial_\nu (\rho_\mu^c - \gamma_5 a_{1\mu}^c)
({\bf F}^c_{(8)})_{ab} N^b_{(8m)} + h.c. \, ,
\\ \nonumber && \left(\frac{\kappa^B_{(8)}}{2M} \right) \overline N^a_{(8)}
\sigma^{\mu\nu} \partial_\nu (\rho_\mu^c - \gamma_5 a_{1\mu}^c)
({\bf D}^c_{(8)})_{ab} N^b_{(8m)} + h.c. \, .
\label{e:8/8m_Pauli_interact}
\end{eqnarray}

\item
For $[(\bf 10,\bf 1)\oplus(\bf 1,\bf 10)]$ - $[(\bf 1,\bf 10)\oplus(\bf
10,\bf 1)]_{\rm [mir]}$ the chiral invariant interactions are:
\begin{eqnarray}
\left(\frac{\kappa_{(10)}}{2M} \right) \overline \Delta^a_{(10)}
\sigma^{\mu\nu} \partial_\nu (\rho_\mu^c - \gamma_5 a_{1\mu}^c)
({\bf F}^c_{(10)})_{ab} \Delta^b_{(10m)} + h.c. \, .
\label{e:10/10m_Pauli_interact}
\end{eqnarray}

\item
For $[({\bf 6},{\bf 3})\oplus({\bf 3},{\bf 6})]$ - $[({\bf 3},
\overline{\bf 3}) \oplus (\overline{\bf 3}, {\bf 3})]$
the chiral invariant interactions are:
\begin{eqnarray}
\left(\frac{\kappa_{(9/18)}}{2M} \right) \overline N^a_{(9)}
\sigma^{\mu\nu} \partial_\nu (\rho_\mu^c + \gamma_5 a_{1\mu}^c)
({\bf T}^c_{(9/18)})_{ab} N^b_{(18)} + h.c. \, .
\label{e:9/18_Pauli_interact}
\end{eqnarray}

\item
For $[({\bf 8},{\bf 1})\oplus({\bf 1},{\bf 8})]$ -
$[({\bf 1},{\bf 10})\oplus({\bf 10},{\bf 1})]_{\rm [mir]}$
the chiral invariant interactions are:
\begin{eqnarray}
\left(\frac{\kappa_{(8/10)}}{2M} \right) \overline N^a_{(8)}
\sigma^{\mu\nu} \partial_\nu (\rho_\mu^c - \gamma_5 a_{1\mu}^c)
({\bf T}^c_{(8/10)})_{ab} N^b_{(10m)} + h.c. \, .
\label{e:8/10m_Pauli_interact}
\end{eqnarray}

\end{enumerate}

\begin{table}[hbt]
\caption{Allowed chiral invariant Pauli type interaction terms
with one $(\mathbf{8},\mathbf{1})\oplus(\mathbf{1},\mathbf{8})$
vector meson field $\bar N \sigma^{\mu\nu} \partial_\nu M_\mu N$.
In the first column we show the chiral representation of $N$, and
the first row the chiral representation of $\bar N$. We use ``[mir]''
to denote the relevant mirror fields.}
\begin{center}
\label{tab:Pauli interactions}
\begin{tabular}{c|c|c|c|c}
\hline \hline &
$(\mathbf{8},\mathbf{1})\oplus(\mathbf{1},\mathbf{8})$ &
$(\mathbf{3},\mathbf{\bar 3})\oplus(\mathbf{\bar 3},\mathbf{3})$ &
$(\mathbf{6},\mathbf{3}) \oplus (\mathbf{3},\mathbf{6})$ &
$(\mathbf{10},\mathbf{1})\oplus(\mathbf{1},\mathbf{10})$ \\
\hline $(\mathbf{1},\mathbf{8})\oplus(\mathbf{8},\mathbf{1})$[mir] &
2 $\times M_\mu$ &  &  & $M_\mu$ \\
\hline $(\mathbf{ 3},\mathbf{\bar 3})\oplus(\mathbf{\bar 3},
\mathbf{3})$[mir] & & $M_\mu$, $M^\dagger_\mu$ & & \\
\hline $(\mathbf{\bar 3},\mathbf{\bar 6}) \oplus (\mathbf{\bar
6},\mathbf{\bar 3})$[mir] & & & $M_\mu$, $M^\dagger_\mu$ &
\\ \hline $(\mathbf{1},\mathbf{\overline{10}})\oplus(\mathbf{\overline{10}},
\mathbf{1})$[mir] & $M_\mu$ & & & $M_\mu$
\\ \hline
\hline & $(\mathbf{3},\mathbf{\bar 3})\oplus(\mathbf{\bar 3},\mathbf{3})$
& $(\mathbf{6},\mathbf{3}) \oplus (\mathbf{3},\mathbf{6})$
\\ \cline{1-3} $(\mathbf{\bar 3},\mathbf{
3})\oplus(\mathbf{3},\mathbf{\bar 3})$ & & $M_\mu$  \\
\cline{1-3} $(\mathbf{\bar 6},\mathbf{\bar 3}) \oplus
(\mathbf{\bar 3},\mathbf{\bar 6})$ & $M^\dagger_\mu$ &
\\ \hline \hline
\end{tabular}
\end{center}
\end{table}

From the above summary of the interactions, we note that for the
normal-normal (also known as the ``naive-naive'') combination, only
$[(\mathbf{6},\mathbf{3})\oplus(\mathbf{3},\mathbf{6})]$ -
$[(\mathbf{3},\mathbf{\bar 3})\oplus(\mathbf{\bar 3},\mathbf{3})]$
survives, whereas the naive-mirror one
$[(\mathbf{6},\mathbf{3})\oplus(\mathbf{3},\mathbf{6})]$ -
$[(\mathbf{\bar 3},\mathbf{3}) \oplus (\mathbf{3},\mathbf{\bar
3})]_{\rm [mir]}$ vanishes,
which is a  selection rule due to $SU(3) \times SU(3)$ chiral symmetry.
Furthermore, as the first mixing term preserves the $U_A(1)$ symmetry,
while the second one does not, we are forced to conclude
that the selection rule leads to the Harari scenario
where the $U_A(1)$ symmetry is maintained as the only viable one
in this three-quark baryon field and no chiral mixing in vector
mesons approximation, whereas the Gerstein-Lee one is effectively
ruled out by the chiral selection rule~\cite{Chen:2010ba}.

In order to have a realistic anomalous magnetic moment it may be
necessary to include the flavor-singlet, chiral-singlet vector
meson $\phi_{\mu}$. Once again, there are chiral selection rules
that strongly prefer the Harari scenario.

\subsection{The Anomalous Magnetic Moment Results: Comparison with
Experiment in the $SU(3)$ Symmetry Limit}
\label{ssect:results}

Thus far we have studied the anomalous magnetic moments of baryon
fields and found specific constraints due to chiral symmetry.
The basic quantity that we address is the $D/F$ ratio for the baryon anomalous
magnetic moments, whose ``experimental value'' has been extrapolated to 
$D/F\simeq$3 in the $SU(3)$ symmetry limit.
Note that that is precisely the value that shows up in the $[({\bf
6},{\bf 3})\oplus({\bf 3},{\bf 6})]$ -- $[({\bf 3}, \overline{\bf
3}) \oplus ( \overline{\bf 3}, {\bf 3})]$ baryon mixing
interaction. Indeed, all of the other chiral interactions have a vanishing
$D$ components.

Of course, it is not one, but a linear combination (``admixture") of
three chiral representations that describe the physical baryon states,
as explained in the Introduction and in Refs~
\cite{Chen:2008qv,Chen:2009sf,Dmitrasinovic:2009vp,Chen:2010ba}
where we found two candidate chiral mixing scenarios: a) the Harari one,
i.e. $[({\bf 6},{\bf 3})\oplus({\bf 3},{\bf 6})]$-$[({\bf 3},
\overline{\bf 3}) \oplus ( \overline{\bf 3}, {\bf
3})]$-$[(\overline{\bf 3},{\bf 3}) \oplus ( {\bf 3},\overline{\bf
3})]$; and b) the Lee-Gerstein one, i.e. $[({\bf 6},{\bf
3})\oplus({\bf 3},{\bf 6})]$-$[({\bf 8},{\bf 1})\oplus ({\bf
1},{\bf 8})]$-$[(\overline{\bf 3},{\bf 3}) \oplus ( {\bf
3},\overline{\bf 3})]$. As the $[({\bf 3}, \overline{\bf 3}) \oplus
(\overline{\bf 3}, {\bf 3})]$ multiplet shows up only in the
Harari scenario, this is a ``smoking gun" evidence
supporting it, and overturning the Gerstein-Lee one,
subject to the no-chiral-mixing assumption in the vector meson sector.

Next we may consider the chiral mixing for the $[({\bf8}, {\bf 1})
\oplus ({\bf 1},{\bf 8})]$ vector mesons  with the
$[(\overline{\bf 3}, {\bf 3}) \oplus ({\bf 3},\overline{\bf 3})]$
component. One may use our old results, Refs.~
\cite{Chen:2010ba,Dmitrasinovic:2000ei,Dmitrasinovic:2000gx}, to do
so and relax this last assumption: the $[(\overline{\bf 3}, {\bf 3}) \oplus
({\bf 3}, \overline{\bf 3})]$ chiral component of the vector
mesons couples magnetically to the baryons chiral multiplets in
exactly the same fashion as the spinless $[(\overline{\bf 3}, {\bf
3}) \oplus ({\bf 3}, \overline{\bf 3})]$ mesons treated in
Ref.~\cite{Chen:2010ba}. So we may use Eq.~(35) in Ref.~\cite{Chen:2010ba}
to read off the $F$ and $D$ values of the anomalous magnetic moments
in such a scheme: they are $F=1$ and $D=0$, thus leading to $D/F = 0$,
again in stark contrast to the ``experimental'' value $D/F\simeq$3 in
the $SU(3)$ symmetry limit. This eliminates the chiral mixing of
vector meson as a viable explanation of the baryons' magnetic moments.
Indeed, it seems to imply certain limits on the amount of such
chiral mixing, that will be explored elsewhere.

Note that these results hold even in the chiral limit and have
nothing to do with the value of the pion-nucleon $\Sigma$-term as
suggested in Ref.~\cite{Donoghue:1985bu}. Moreover, the
chiral/flavour-singlet vector meson field couples with arbitrary
strength to baryons, which introduces arbitrary ``strange''
anomalous magnetic moment, again even in the chiral limit.

\section{Summary and Conclusions}
\label{sect:summary}

We have used the results of our previous papers~\cite{Chen:2008qv,Chen:2009sf}
to construct the $SU_L(3) \times SU_R(3)$ chiral invariant
interactions of baryon fields with vector mesons.
This approach is based on the chiral
$[({\bf 6},{\bf 3})\oplus({\bf 3},{\bf 6})]$ multiplet
mixing with the chiral
$[(\mathbf{3},\mathbf{\bar 3}) \oplus (\mathbf{ \bar
3},\mathbf{3})]$ and $[(\mathbf{ 8},\mathbf{1}) \oplus
(\mathbf{1},\mathbf{8})]$ multiplets and
is constrained by the well known phenomenological facts
regarding the baryon axial currents.

The results of the three-field (``two-angle'') mixing were ambiguous
insofar as all phenomenologically permissible combinations of
interpolating fields lead to the same $F$,$D$ values, in reasonable
agreement with the result extrapolated from experiment in the $SU(3)$ symmetry limit.
That led to two permissible scenarios: a) the Gerstein-Lee~\cite{Gerstein:1966zz},
and b) the Harari scenario~\cite{Harari:1966yq,Harari:1966jz}, neither of
which could be eliminated on the basis of axial currents and baryon
masses alone.

What was left unfinished were the magnetic moments of the baryon octet.
Here we attacked that problem by first constructing all
$SU_L(3) \times SU_R(3)$ chirally symmetric
baryon-one-vector-meson interactions that mix the three basic baryon
chiral multiplets (and their mirror counterparts). All of these
chiral interactions obey the $U_{A}(1)$ symmetry, as well.

We used the resulting interactions' chiral selection rules to
select the only scenario that can reproduce the observed anomalous
magnetic moments: the Harari scenario.
Moreover, the magnetic moment $F/D \simeq 1/3$ predicted by the
chiral interaction, has the same value as in the $SU(6_{\rm FS})$
symmetry limit, or as in the  non-relativistic quark model.
This last fact is curious and requires further investigation.

The next step, left for the future, is to investigate the $SU_L(3)
\times SU_R(3) \rightarrow SU_L(2) \times SU_R(2)$ symmetry
breaking and the study of the chiral $SU_L(2) \times SU_R(2)$
properties of hyperons. Then one may consider explicit chiral
symmetry breaking corrections to the axial and the vector
currents, which are related to the $SU_L(3) \times SU_R(3)$
symmetry breaking meson-nucleon derivative interactions, not just
the explicit $SU(3)$ symmetry breaking ones that have been
considered thus far (see Ref.~\cite{Yamanishi:2007zza} and the
previous subsection, above).

We finish on a historical note: even though chiral
mixing has been known for more than 40 years
~\cite{Hara:1965,Lee:1968,Bardeen:1969ra,Weinberg:1969hw},
the $SU_L(3) \times SU_R(3)$ chiral interactions necessary to
describe the anomalous magnetic moments have not been discussed in
print, only the problems associated with them~\cite{Dashen:1966zz}.
Moreover, it ought to be noted that Gerstein and Lee~\cite{Gerstein:1966zy}
had calculated anomalous magnetic moments of the nucleons that were in
agreement with experiment in their chiral mixing scheme.
These authors apparently did not try to extend their scheme to
hyperons, however, nor did they construct a chiral Lagrangian
that reproduces such chiral mixing.

\section*{Acknowledgments}
\label{ack}

We wish to thank Profs. Daisuke Jido and Hiroshi Toki for insisting on the completion of this work. One of us (V.D.) wishes to thank the RCNP, Osaka University and the Yukawa Institute for Theoretical Physics, Kyoto, under whose auspices this work was begun. The work of H.X.C is supported by the National Natural Science Foundation of China under Grant No. 11147140, and the Scientific Research Foundation for the Returned Overseas Chinese Scholars, State Education Ministry. The work of V.D. was supported by the Serbian Ministry of Science and Technological Development under grant number 141025. The work of A.H. is partly supported by the Grant-in-Aid for Scientific Research on Priority Areas ``Elucidation of New Hadrons with a Variety of Flavors (E01: 21105006)'' from the ministry of Education, Culture, Sports, Science and Technology of Japan.

\end{document}